# Resistive switching phenomena in TiO$_x$ nanoparticle layers for memory applications

Emanuelle Goren,[1] Mariana Ungureanu,[2] Raul Zazpe,[2] Marcelo Rozenberg,[3,4] Luis E. Hueso[2,5], Pablo Stoliar,[2,6,*] Yoed Tsur,[1] and Fèlix Casanova[2,5]

[1]RBNI and Chemical Engineering Department, Technion, 3200003 Haifa, Israel
[2]CIC nanoGUNE, 20018 Donostia-San Sebastian, Basque Country, Spain
[3]Laboratoire de Physique des Solides, CNRS UMR 8502, Université Paris Sud, Bât 510, 91405 Orsay, France
[4]Departamento de Física and IFIBA-Conicet, FCEyN, UBA, Ciudad Universitaria, P.1, Buenos Aires 1428, Argentina
[5]IKERBASQUE, Basque Foundation for Science, 48011 Bilbao, Basque Country, Spain
[6]ECyT, Universidad Nacional de San Martín, Campus Miguelete, 1650 San Martín, Argentina
* pstoliar@nanogune.eu

**Abstract**

Electrical characteristics of a Co/TiO$_x$/Co resistive memory device, fabricated by two different methods are reported. In addition to crystalline TiO$_2$ layers fabricated via conventional atomic layer deposition (ALD), an alternative method has been examined, where TiO$_x$ nanoparticle layers were fabricated via sol-gel. The different devices have shown different hysteresis loops with a unique crossing point for the sol–gel devices. A simple qualitative model is introduced to describe the different current-voltage behaviours by suggesting only one active metal-oxide interface for the ALD devices and two active metal-oxide interfaces for the sol-gel devices. Furthermore, we show that the resistive switching behaviour could be easily tuned by proper interface engineering and that despite having a similar active material, different fabrication methods can lead to dissimilar resistive switching properties.

Resistive switching (RS) is emerging as a promising mechanism for future generation of memory devices.[1-3] It refers to the reversible change in the resistivity of a dielectric media by the application of electric pulses.[4] The typical RS memory cell is composed of a thin oxide film that is sandwiched between two metallic electrodes. Electric pulses applied between these electrodes switch the resistance state of the oxide between two non-volatile states. Among the many binary and complex oxides already proposed as active layer in these RS devices, $TiO_2$ (grown by sputtering, atomic layer deposition (ALD) or thermal evaporation) is one of the most often used materials.[5-10] Motivated by the possibility of scaling down to nanomaterials, which would lead to reduced fabrication costs of such nanodevices, we investigated the RS phenomenon in $TiO_x$ nanoparticle layers. Up to now, only few reports have been published for resistive switching in nanoparticle layers, leaving unanswered questions regarding the conduction mechanisms occurring in these peculiar systems.[11-15]

In this letter, we study the resistive switching phenomena in devices with $TiO_x$ nanoparticles (spheres of about 3 nm in diameter) as active layer, fabricated using sol-gel technique and compare it to devices using conventional $TiO_2$ thin films prepared by ALD. We discuss the main differences between the resistive switching phenomena observed in each case and propose a model showing that these differences arise from the fact that opposite interfaces govern the resistive switching in each one. Whereas in the ALD-based devices the bottom interface is not switching, in the sol-gel-based devices the bottom interface shows a larger resistance modulation. Our results will contribute to the incorporation of nanoparticles into real memory devices of reduced dimensions at lower costs.

The studied devices are vertical stacks of metal/$TiO_x$/metal deposited on a $SiO_2$/Si wafer (Fig. 1(a)). The fabrication involves first the deposition of a 40-nm-thick Co bottom electrode on the $SiO_2$/Si substrate. Next, a 55-nm-thick $TiO_x$ layer is grown as described below, and finally Co (40 nm thick) / Pd (35 nm thick) square top contacts with 100 μm x 100 μm area are defined by photolithography. We have two different types of samples: samples with the TiO$_2$ thin film prepared by ALD [16] and samples with the TiO$_x$ layer prepared by sol-gel method. [17]

In the case of the ALD-based devices, TiO$_2$ films were deposited by ALD on top of the Co bottom electrode using Tetrakis(dimethylamino)titanium as the precursor, H$_2$O vapor as the oxidant, and N$_2$ as the purging gas between pulses. The films were grown at 210 °C. The purge time between the Ti precursor and the H$_2$O was 5 s, and between the H$_2$O and a second pulse of Ti precursor was 10 s. The film is crystalline, as seen in the High Resolution Transmission Electron Microscopy (HRTEM) images (Figs. 1(b) and 1(c)). For the sol-gel-based devices, TiO$_x$ was prepared as follows.[17] Titanium isopropoxide Ti[OCH(CH$_3$)$_2$]$_4$ (97% 4.5 ml from Aldrich) was used as a precursor. The precursor was mixed with n-propanol CH$_3$CH$_2$CH$_2$OH (99.5%, 10 ml from Aldrich) and AcAc C$_5$H$_8$O$_2$ (99%, 1.6 ml from Aldrich). Afterwards the solution was mixed and 1 ml was taken from it into a new vial where it was mixed with n-propanol CH$_3$CH$_2$CH$_2$OH (99.5%, 1.6 ml from Aldrich) and 0.2 ml of de-ionized water purified using a milli-Q system. The TiO$_x$ solution was spin coated directly on the Co bottom electrode at 8000 RPM. The sample was annealed at 260 °C for 2 h in air. The layer is composed of amorphous nanoparticles of about 3 nm diameter, according to HRTEM images (Figs. 1(d) and 1(e)).

Figure 2 shows the electrical characterization of our devices, which is based on current-voltage (I-V) curves and hysteresis-switching loops (HSLs), [18, 19] both performed at room temperature with the top and bottom electrodes biased and grounded respectively. The I-V curves in Figs. 2(a) and 2(b) were obtained by sweeping the applied voltage, *V*, with a staircase sequence 0 V → +5 V → –5 V → 0 V in steps of 0.1 V. Figure 2(a) shows that the ALD-based devices, initially in low resistance state (LRS), are switched to high resistance state (HRS) during the positive branch of the sweep; i.e. when positive voltage is applied to the top contact. The negative part of the sweep drives the system back to LRS. In contrast, in Fig. 2(b) we see that the sol-gel-based devices, initially in HRS, are switched to LRS when positive voltage is applied to the Co/Pd top contact. At negative voltage, this device switches from HRS to LRS; yet, we observe an extra crossing point at about −1 V. Both ALD-based devices and sol-gel-based devices present bipolar resistive switching, which means that electric pulses of alternate polarity are needed in order to reversibly switch the state from LRS to HRS and back. [3]

The HSLs presented in Figs. 2(c) and 2(d) were obtained by applying a sequence of writing voltage pulses, $V_{write}$, that follows 0 V → +5 V → −5 V → 0 V in steps of 0.5 V. After each pulse we waited 20 ms at 0 V in order to discharge the capacitive effects and then we measured the remnant current, $I_{rem}$, with a fixed reading voltage of +2 V. [18] The +2 V value was chosen considering that around this voltage we have the highest ratio between the LRS and HRS on the I-V curves. These HSLs reflect a non-volatile resistive memory behavior. We observe a clockwise loop for the ALD-based device (see Fig. 2(c)). It requires a minimum voltage of +3 V to be RESET (that is, the transition from LRS to HRS) and a $V_{write}$ < -1 V for the SET (switch from HRS to LRS). On the contrary, the sol-gel-based device shows a counter clockwise loop without a clear threshold for the SET and a minimum RESET voltage of about –2 V (Fig. 2(d)).

These HSLs confirm that our devices present bipolar resistive switching. The mechanism causing bipolar RS in $TiO_x$ samples is generally electric-field-driven modulation of the concentration of oxygen vacancies at the interfaces with the metallic electrodes. [19-22] In particular, the HSL of the ALD-based device (Fig. 2(c)) shows a resistance modulation of factor 10 that can be simply rationalized as follows. As mentioned, when positive pulses exceed +3 V, the whole resistance of the device increases (i.e., $I_{rem}$ decreases). Positive pulses might either remove oxygen vacancies from the interface with the top electrode (that will increase the resistance of the interface) or increase their concentration in the proximity of the bottom electrode (decreasing the resistance of the interface). At this point we exclude the simultaneous occurrence of both effects (see below). In this case, the observed increase in the resistance with positive polarity indicates that the modulation of vacancies at the interface with the top electrode is the main switching mechanism in the ALD-based samples. With equivalent arguments, we conclude that the control of the oxygen vacancies in the proximity with the bottom electrode is the main switching mechanism in the sol-gel-based samples.

Whereas the I-V curve of the ALD-based device (Fig. 2(a)) is in line with this description, the crossing point at -1 V in the I-V characteristics of the sol-gel-based device (Fig. 2(b)) reveals interplay between both interfaces. In this case, both interfaces are switching, giving rise to this peculiar feature.[22, 23] Above -1 V, the I-V curve is what is expected when only the interface with the bottom electrode switches. For voltages < -1 V, the modulation of the top interface effectively overcomes the effect of the bottom interface. The asymmetry of the I-V curve in the sol-gel-based devices (in the sense that the crossing point is not at zero voltage and that the area enclosed by the I-V curve is larger for positive voltages) indicates that the modulation of the resistance at the bottom interface is significantly larger than in the vicinity of the top electrode. The HSL in Fig. 2(d) only reveals the resistive switching of the bottom interface, because the modulation of the top interface is completely hindered by the reading voltage, i.e. each reading voltage (+2 V) is actually switching the top interface.

We introduce a model in order to rationalize the nontrivial interaction between the interfaces. The basic hypotheses for the model are: i) $TiO_x$ layer becomes conductive due to the presence of oxygen vacancies, ii) interfaces are the more resistive regions of the device and usually one of these interfaces dominates the two-terminal resistance of the device, and iii) voltage pulses result in localized electric fields that drive exchange of ions between the interfaces and the bulk. Under these hypotheses we can deduce, according to the dependencies of the hysteresis with polarity, which of the two interfaces is responsible for the RS effect. As we shall see, the fact that ALD-based and sol-gel-based devices present different behaviour implies that different interfaces are involved in each case.

In this model, the device is divided in three regions: two interface regions where the local density of oxygen vacancies ($[V_O^{\bullet\bullet}]^{t,b}$) modulates the resistance and a bulk region. We use here the Kröger-Vink notation for point defects and add super-indexes *t* and *b* that stand for top and bottom interface. The oxide at the interfaces is expected to remain slightly sub-stoichiometric in any case. The total resistance of the device, *R*, when $[V_O^{\bullet\bullet}]^{t,b} \ll 1$ is given by:

$$R = \left(1 - [V_O^{\bullet\bullet}]^t\right) A^t + R_{bulk} + \left(1 - [V_O^{\bullet\bullet}]^b\right) A^b, \tag{1}$$

where $R_{bulk}$ is the resistance of the bulk (see Fig. 3(a)).[18] At the interfaces, we assume a linear relationship between the density of oxygen vacancies and the resistance, given by the proportionality factors $A^{t,b}$.[22] With this linearization, we are indeed modelling the resistance of the slightly sub-stoichiometric TiO$_x$: the increase of $[V_O^{\bullet\bullet}]^{t,b}$ moves the system away from the stoichiometric condition ($[V_O^{\bullet\bullet}]^{t,b} = 0$), decreasing its resistance.[24] When voltage pulses *V* are applied to the devices, the intensity of the local electric fields developed at the interfaces is:

$$E^{t,b} = V w^{-1} R^{-1} A^{t,b} (1 - [V_O^{\bullet\bullet}]^{t,b}), \tag{2}$$

with *w* being the effective width of the interfaces. These electric fields are responsible for the re-localization of oxygen vacancies at the interfaces that, in turn, modulate their resistance. Indeed, the voltage drop at the interfaces effectively exerts a net force on the oxygen vacancies. Strong enough electric fields overcome the anchoring energy of the ions to the TiO$_x$ matrix and establish an exchange between the interfaces and the bulk.[22] We then propose a drift-diffusion equation for the exchange of oxygen vacancies between the bulk and the interfaces:

$$d[V_O^{\bullet\bullet}]^{t,b}/dt = -\left([V_O^{\bullet\bullet}]^{t,b} - [V_O^{\bullet\bullet}]_0^{t,b}\right) k^{t,b} + \text{sgn}(V) \exp(-E_0^{t,b} + |E^{t,b}|). \tag{3}$$

Here, the first (negative) term stands for the diffusion of vacancies back to the rest condition $\left[V_O^{\bullet\bullet}\right]^{t,b}$, whereas the second term stands for the electric-field-activated drift of ions. $k^{t,b}$ is the diffusion coefficient [25] and $E_0^{t,b}$ the anchoring energy of the oxygen vacancies to the matrix ($E_0^{t,b}$ absorbs any prefactor). This equation describes the following behaviour. When the applied voltage is high enough, the first (diffusion) term can be neglected. The density of vacancies increases or decreases (depending on the voltage polarity, sgn($V$)) at a rate that grows exponentially with the magnitude of the local electric field. At zero or very low voltage, the drift term becomes negligible, and the density of vacancies diffuses to the equilibrium condition $\left[V_O^{\bullet\bullet}\right] = \left[V_O^{\bullet\bullet}\right]_0$ with constant rate $k$.

The numerical outputs based on equations (1)-(3) for two different sets of parameters are presented in Figs. 3(b) and 3(c). In the case of the ALD-based devices (Fig. 3(b)), the simulation qualitatively matches the experiments when only the top electrode switches. This is done by setting $A^b = 0$ and therefore absorbing the (constant) resistance of the bottom interface in $R_{\text{bulk}}$. Basically, any set of parameters ($A^t$, $E_0^t$, $w$ and $R_{\text{bulk}}$) reproduces the general trend and no fine-tuning is needed for this qualitative matching. We also fix $k^b = k^t = 0$, so that no relaxation is considered. In general, we look for the minimum set of parameters needed to reproduce the results. Since the relaxation terms are not necessary to reproduce the results, they are excluded.

Instead, modulation of the oxygen density in both electrodes has to be considered when it comes to simulate the behaviour of the sol-gel-based devices. In fact, we reproduce the qualitative behaviour of the sol-gel-based devices by adding two ingredients as compared to the model for the ALD-based devices: relaxation at the top interface ($k^t$) and resistive switching at the bottom interface ($A^b$, $E_0^b$). Indeed, these parameters introduce the presence of the crossing point in the negative branch of the I-V curve and reproduce the proper direction of the loop (counter clockwise in the positive branch).[26] Here, we also look for the minimum set of parameters in the model and the introduction of these two extra ingredients is the minimum requirement to reproduce the results, while relaxation at the bottom interface is not needed ($k^b = 0$). Figure 3(c) shows a simulation that qualitatively matches the experiments when the modulation of the resistance of this bottom interface is 40% with respect to the top interface ($A^b = 0.4 \cdot A^t$). More importantly, the position of the crossing point in the negative branch is tuned by the anchoring energy of the ions to the matrix. In the simulation plotted in Fig. 3(c), this anchoring energy is lower in the bottom interface than in the top one, $E_0^b/E_0^t = 0.6$.

As can be seen from the HRTEM cross section image of the ALD-based thin film (Fig. 1(b)), a blurred intermediate layer appears at the bottom interface. This interfacial layer is composed of Ti, Co and O and therefore this interface could behave as a lower energy barrier. This is in contrast to the sharp bottom interface seen in the HRTEM cross section image of the sol-gel-based layer (Fig. 1(d)). The different fabrication process might thus account for the different behaviour of the bottom interface, and, in turn, on the RS mechanism. This result suggests that the RS behaviour could be easily tuned by proper interface engineering. We prepared an additional sample in order to experimentally verify this point. The sample contains a sol-gel $TiO_x$ layer (as for the sol-gel devices), but the interface between the Co bottom electrode and this sol-gel layer was modified by the introduction of an intermediate 5-nm-thick layer of $TiO_2$ prepared by ALD, to protect the bottom electrode from the annealing process. The result is a Co/$TiO_2$(ALD, 5 nm)/$TiO_x$(sol-gel, 55 nm)/Co/Pd device. Figure 4 shows the electrical characterization for this sample. The I-V curve (Fig. 4(a)) is clearly distinct from that of the original sol-gel-based device (Fig. 2(b)). The slopes of the I-V curves at positive applied voltages in the HRS are significantly different for both the sol-gel devices, while at negative voltages the crossing point, originally at about -1 V, is now at about -4 V. The key qualitative features of the HSLs in Fig. 4(b) remain unchanged after the modification of the interface with the bottom electrode, when compared to Fig. 2(d): it is also a counter clockwise loop and the SET threshold is not well defined. As predicted, to reproduce this new set of curves we just need to tune the anchoring energy of the oxygen ions in the model proposed for the full sol-gel-based devices. The new value to shift the crossing point to -4 V is $E_0^b/E_0^t = 0.8$. This additional sample supports the bulk-interface-exchange model in our system.

In conclusion, we investigated and compared the resistive switching phenomenon in $TiO_x$ nanoparticle layers prepared by sol-gel, and $TiO_2$ thin films prepared by ALD. Despite having a similar active material, $TiO_x$, different fabrication methods translate into different behaviour of the interfaces, which leads to dissimilar resistive switching properties as evidenced in the I-V curves and hysteresis switching loops. We introduce a simple qualitative model that allows us to rationalize the observed differences. The resistive switching essentially occurs at the top electrode interface in the ALD-based devices, while the bottom interface also plays a role in the sol-gel-based devices. The different fabrication process accounts for the different behaviour of the bottom interface, which can be tuned by proper interface engineering. The observation of resistive switching in nanoparticle layers prepared by sol-gel opens new possibilities for the fabrication of downscaled memory devices at lower costs.

**Acknowledgements**

Work supported by the Spanish Ministry of Economy through Juan de la Cierva (JCI-2010-07837) and Ramon y Cajal (RYC-2012-01031) programs; by the Russell Berrie Nanotechnology Institute at the Technion, Israel's Ministry of Infrastructure, Energy and Water (880002), and Grand Technion Energy Program (GTEP). We thank Dr. S. Baltianski for technical support, ALD was done by Dr. B. Meyler at the Technion's Micro and Nano Electronics Centre and HRTEM images were taken by Dr. Y. Kauffmann, Department of Materials Science & Engineering, Technion.

**References**

[1] A. Sawa, Mater. Today **11**, 28–36 (2008).

[2] D. S. Jeong, R. Thomas, R. S. Katiyar, J. F. Scott, H. Kohlstedt, A. Petraru, and C. S. Hwang, Rep. Prog. Phys. **75**, 076502 (2012).

[3] R. Waser, and M. Aono, Nature Mater. **6**, 833–840 (2007).

[4] M. Rozenberg, Scholarpedia J. **6**, 11414 (2011).

[5] B. J. Choi, D. S. Jeong, S. K. Kim, C. Rohde, S. Choi, J. H. Oh, H. J. Kim, C. S. Hwang, K. Szot, R. Waser, B. Reichenberg, and S. Tiedke, J. Appl. Phys. **98**, 033715 (2005).

[6] D. H. Kwon, K. M. Kim, J. H. Jang, J. M. Jeon, M. H. Lee, G. H. Kim, X.-S. Li, G.-S. Park, B. Lee, S. Han, M. Kim, and C. S. Hwang, Nature Nanotechnology **5**, 148 (2010).

[7] D. S. Jeong, H. Schroeder, and R. Waser, Electrochemical and solid-state letters **10**, G51 (2007).

[8] C. Yoshida, K. Tsunoda, H. Noshiro, and Y. Sugiyama, Appl. Phys. Lett. **91**, 223510 (2007).

[9] J. J. Yang, M. D. Pickett, X. Li, D. AA Ohlberg, D. R. Stewart, and R. S. Williams, Nature nanotechnology **3**, 429 (2008).

[10] N. Ghenzi, D. Rubi. E. Mangano, G. Gimenez, J. Lell, A. Zelcer, P. Stoliar, and P. Levy, Thin Solid Films **550**, 683-688 (2014).

[11] N. Gergel-Hackett, B. Hamadani, B. Dunlap, J. Suehle, C. Richter, C. Hacker, and D. Gundlach, IEEE Electron Device Letters **30**, 706-708 (2009).

[12] C. Lee, I. Kim, W. Choi, H. Shin, and J. Cho, Langmuir **25**, 4274-4278 (2009).

[13] N. Ghenzi, P. Stoliar, M.C. Fuertes, F. G. Marlasca, and P. Levy, Physica B **407**, 3096-3098 (2012).

[14] J. Yun, K. Cho, B. Park, B. H. Park, and S. Kim, J. Mater. Chem. **19**, 2082-2085 (2009).

[15] K. P. Biju, X. J. Liu, E. M. Bourim, I. Kim, S. Jung, M. Siddik, J. Lee, and H. Hwang, J. Phys. D: Appl. Phys. **43**, 495104 (2010).

[16] S. M. George, Chem. Rev. **110**, 111-131 (2010).

[17] Y. Paz, Z. Luo, L. Rabenberg, and A. Heller, J. Mater. Res. **10**, 2842 (1995).

[18] R. Zazpe, P. Stoliar, F. Golmar, R. Llopis, F. Casanova, and L. E. Hueso, Appl. Phys. Lett. **103**, 073114 (2013).

[19] J. J. Yang, I. H. Inoue, T. Mikolajick, and C. S. Hwang, MRS Bull. **37**, 131 (2012).

[20] D. B. Strukov, G. S. Snider, D. R. Stewart, and R. S. Williams, Nature **453**, 80 (2008).

[21] H. Y. Jeong, J. Y. Lee, and S.-Y. Choi, Appl. Phys. Lett. **97**, 042109 (2010).

[22] M. J. Rozenberg, M. J. Sánchez, R. Weht, C. Acha, F. Gomez-Marlasca, and P. Levy, Phys. Rev. B **81**, 115101 (2010).

[23] M. J. Rozenberg, I. H. Inoue and M. J. Sánchez, Phys. Rev. Lett. **92**, 178302 (2004).

[24] N. Ghenzi, M. J. Sánchez, and Pablo Levy, J. Phys. D: Appl. Phys. **46**, 415101 (2013).

[25] F. Assad, K. Banoo, and M. Lundstrom, Solid-State Electronics **42**, 283-295 (1998).

[26] Values of the relaxation parameter $k^t$ between 10 and 40, expressed in units of the numerical integration time step, qualitatively reproduce the I-V curve of Fig. 2(b). The number of steps in these simulated I-V curves is the same as the number of voltage steps in Fig. 2 (b) during a complete cycle. Since the voltage step time used in the experiment is 60 ms, $k^t = 10$ corresponds to a decay constant of 600 ms. In order to have a direct measurement of the retention time of the top interface, we measured HSLs with different waiting times after each writing pulse (between 20 ms and 5 s). Then, by fitting the HSL current values at $V_{write} = 0$ with the exponential decay equation $I_{rem}(0^+) / I_{rem}(0^-) - 1 = A_\infty + A_1 \exp(-t/\tau)$, we extract a decay constant of $\tau \approx 100$ ms. which is in very good agreement with the one obtained from the numerical simulations of our toy model.

**Figures**

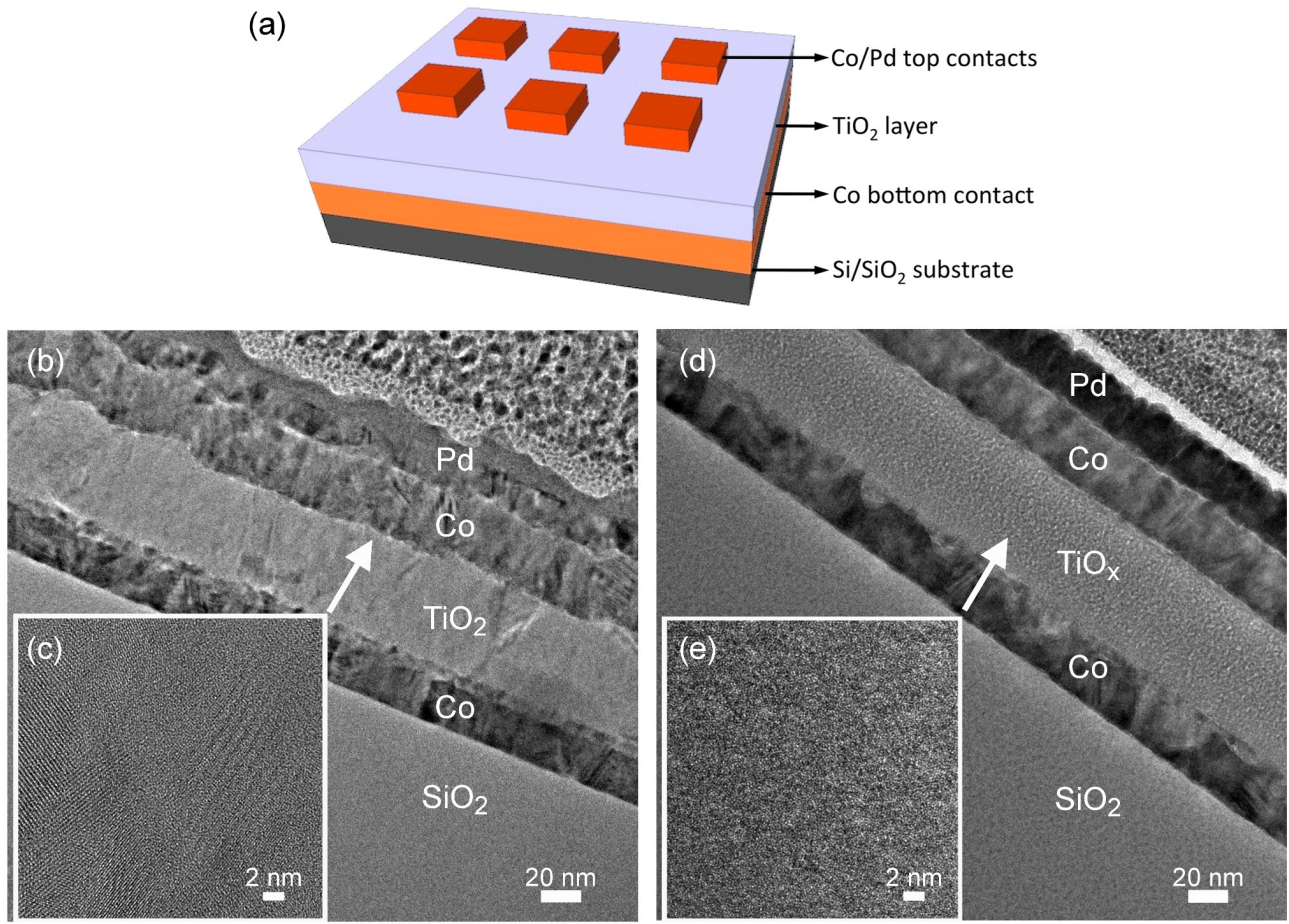

**FIG. 1.** (a) Schematic representation of the studied devices. A Co film is deposited on the SiO$_2$/Si substrate to serve as bottom electrode. Then a TiO$_x$ layer is deposited (either by ALD or by sol-gel) followed by Co/Pd top contacts to serve as top electrodes. (b-e) High Resolution Transmission Electron Microscopy images. (b) Cross section of an ALD-based device. (c) Zoom-in of the TiO$_2$ film prepared by ALD showing the crystalline nature of the film. (d) Cross section of a sol-gel-based device. (e) Zoom-in of the TiO$_x$ film prepared by sol-gel showing the amorphous nature of the nanoparticles forming the layer.

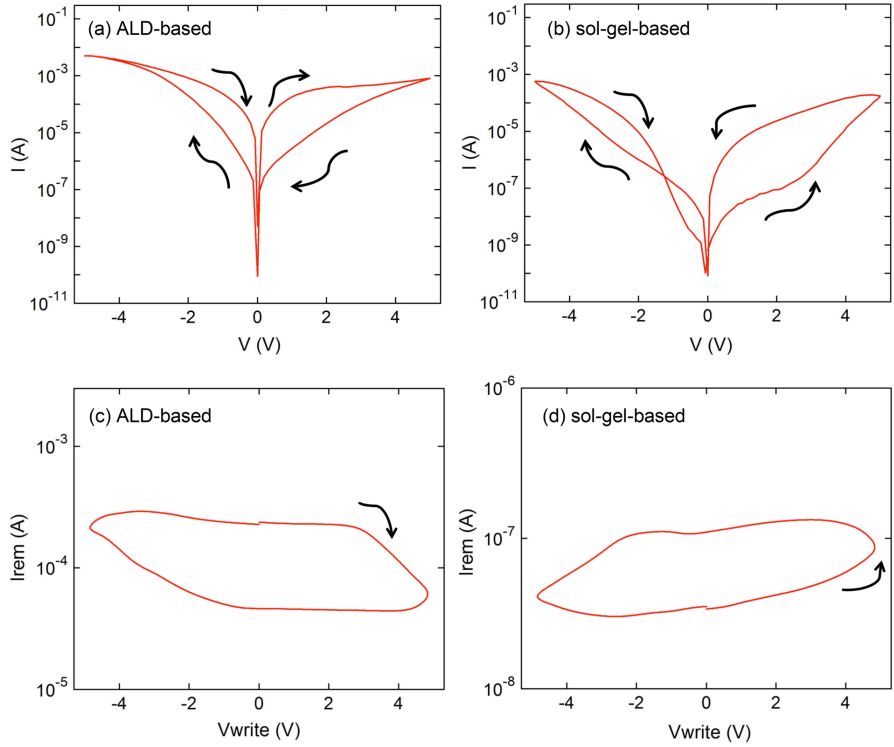

**FIG. 2.** I–V characteristics for (a) an ALD-based device and (b) a sol-gel-based device. The current is plotted as the logarithm of the absolute value. The arrows indicate the path of the applied voltage sweep. Remnant current ($I_{rem}$) hysteresis switching loops for (c) the ALD-based device and (d) the sol-gel-based device at a reading voltage of +2 V after we apply a writing voltage pulse ($V_{write}$) immediately followed by a waiting time of 20 ms at 0 V to discharge capacitive effects. The arrows indicate to direction of the loop when sweeping the writing voltage.

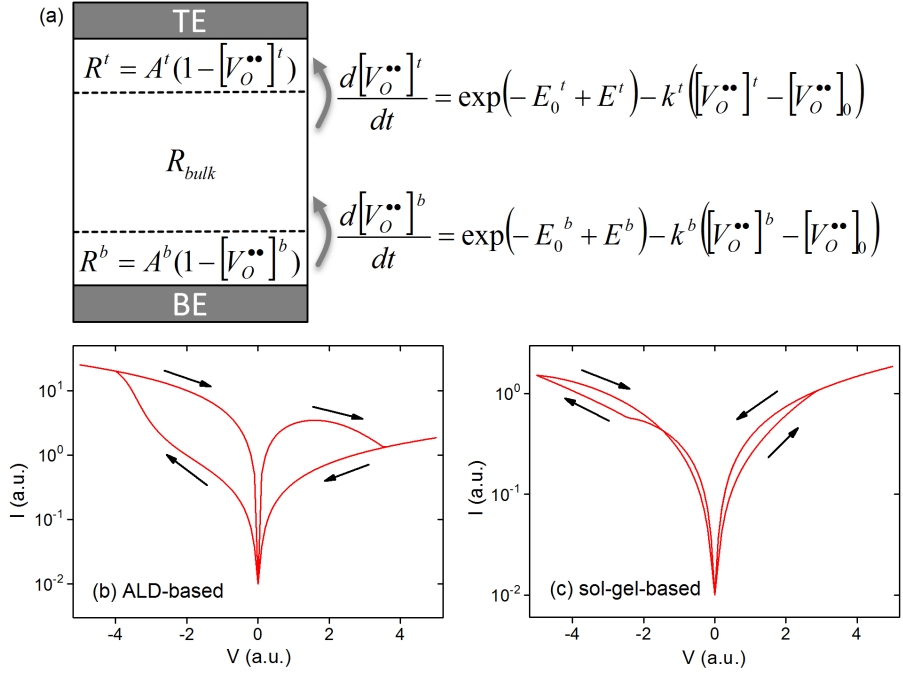

FIG. 3. (a) Scheme of the proposed model for the resistive switching mechanism. In the ALD-based devices the movement of oxygen ions only occurs in the proximity of the top electrode, whereas in the sol-gel-based devices occurs in both interfaces. Numerical simulations showing a qualitative matching with the experimental I-V curves for (b) the ALD-based devices and (c) the sol-gel-based devices. The arrows indicate the path of the applied voltage sweep.

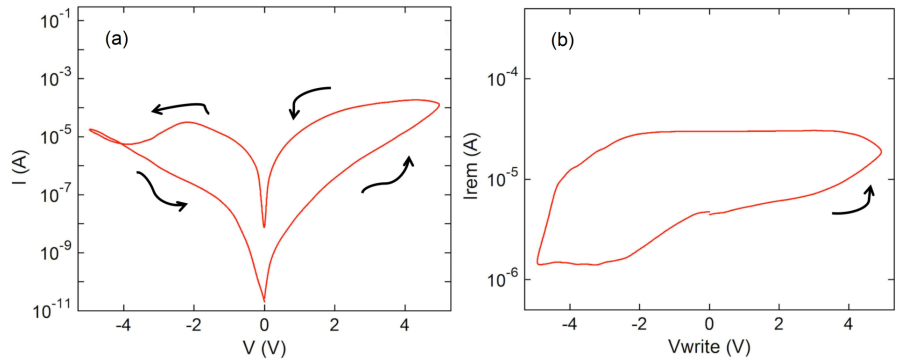

**FIG. 4.** (a) I–V characteristics for a sol-gel-based device in which the bottom interface have been modified by first depositing 5 nm of $TiO_2$ by ALD. The arrows indicate the path of the applied voltage sweep. (b) Remnant current ($I_{rem}$) hysteresis switching loop measured at the same device at a reading voltage of +2 V after we apply a writing voltage pulse ($V_{write}$) immediately followed by a waiting time of 20 ms at 0 V to discharge capacitive effects. The arrows indicate to direction of the loop when sweeping the writing voltage.